\documentclass[12pt]{iopart}

\usepackage{graphicx}

\usepackage{iopams}

\usepackage{rotating}

\newtheorem{theorem}{Theorem}

\newtheorem{corollary}{Corollary}

\newtheorem{proposition}{Proposition}

\begin{document}

\title[ ]
{Internal labelling operators and contractions of Lie algebras}

\author{R. Campoamor-Stursberg\dag}

\address{\dag\ Dpto. Geometr\'{\i}a y Topolog\'{\i}a\\Fac. CC. Matem\'aticas\\
Universidad Complutense de Madrid\\Plaza de Ciencias, 3\\E-28040
Madrid, Spain}

\ead{rutwig@mat.ucm.es}

\begin{abstract}
We analyze under which conditions the missing label problem
associated to a reduction chain $\frak{s}^{\prime}\subset
\frak{s}$ of (simple) Lie algebras can be completely solved by
means of an In\"on\"u-Wigner contraction $\frak{g}$ naturally
related to the embedding. This provides a new interpretation of
the missing label operators in terms of the Casimir operators of
the contracted algebra, and shows that the available labeling
operators are not completely equivalent. Further, the procedure is
used to obtain upper bounds for the number of invariants of affine
Lie algebras arising as contractions of semisimple algebras.

\end{abstract}


\maketitle



\section{Introduction}

\medskip

A recurring problem in group theoretical applications to physical
problems is the reduction of irreducible representations of a Lie
group into multiplets of some subgroup of internal symmetry.
Sometimes, and depending on the nature of the embedding, the
subgroup does not provide enough labels to distinguish the basis
states without ambiguity. We are therefore led to find additional
operators to separate those states not properly described by the
subgroup labels. Various techniques have been developed to
surmount this difficulty, such as the projection technique of
Elliott for the reduction chain $\frak{su}(3)\supset \frak{so}(3)$
used in atomic physics, the method of elementary multiplets in the
spectroscopic chain $\frak{so}(7)\supset G_{2}\supset\frak{so}(3)$
 to describe $f$ electron configurations of rare earths, or the
construction of integrity bases in the enveloping algebras for the
Wigner supermultiplet model $\frak{su}(4)\supset
\frak{su}(2)\times\frak{su}(2)$, among others \cite{El}. More
recently, K-matrix theory and the rotor expansion method have been
shown to be powerful techniques to solve the missing label problem
in many important problems, like the nuclear $\frak{sp}(3)$ model
\cite{Ro,Ro2}.

A complementary analytical approach to the so-called missing label
problem (MLP) was developed in \cite{Sh2,Sh}, by means of basis
functions that are common eigenstates of commuting operators. This
point of view also allows to recover the missing operators as
subgroup scalars in the enveloping algebra of $\frak{s}$, as well
as to compute them as solutions of a system of partial
differential equations. Although this approach has been the less
used for solving the MLP, it presents some interesting features
over the pure algebraic method of enveloping algebras. It has been
observed in the literature that symmetry breaking is, to some
extent, equivalent to consider contractions of Lie algebras
\cite{Ce}. In this sense, the symmetry preserved corresponds to
some subalgebra which remains unchanged by the contraction. At
least for the $\frak{su}(3)$ model, this idea has been developed
by means of the rotor expansion \cite{Ro2}.

This is the point of view we adopt in this work. More
specifically, we combine the analytical method of \cite{Sh} for
solving the MLP with contractions of Lie algebras. We prove that
for any embedding $\frak{s}\supset\frak{s}^{\prime}$ of
(semisimple) Lie algebras, there is an associated simple
In\"on\"u-Wigner contraction of $\frak{s}$ onto an affine Lie
algebra
$\frak{g}=\frak{s}^{\prime}\overrightarrow{\oplus}_{R}nL_{1}$,
where $nL_{1}$ denotes an $n$-dimensional Abelian algebra and $R$
is a representation of the subalgebra $\frak{s}^{\prime}$ such
that the adjoint representation $ad$ of $\frak{s}$ satisfies the
condition $ad(\frak{s})=ad(\frak{s}^{\prime})\oplus R$. It is
further proven that any invariant of the contraction $\frak{g}$
can be formally taken as missing label operator. It is therefore
reasonable to study whether the invariants of the contraction
$\frak{g}$ are sufficient in number to provide a set of missing
label operators, and therefore, to completely solve the missing
label problem. We characterize when it is possible to solve the
MLP by means of this associated contraction, and derive some
useful consequences for the number of invariants of inhomogeneous
Lie algebras. One important fact arises from this method, namely,
that the missing label operators obtained inherit an intrinsic
meaning as terms of invariants that disappear during contraction,
and should correspond to the natural choice of operators, since
they are internally determined by the group-subgroup chain. For
the case of no missing labels, we extract an interesting
consequence, namely, that the invariants of the contraction arise
as polynomial functions of the Casimir operators of the contracted
Lie algebra $\frak{s}$ and the subalgebra $\frak{s}^{\prime}$.
This enables us to determine upper bounds for the number of
inhomogeneous Lie algebras that appear as contractions of
semisimple Lie algebras.

\medskip

It is known from the classical theory that irreducible
representations of semisimple Lie algebras are labelled
unambigously by the eigenvalues of Casimir operators. More
generally, it can be established that irreducible representations
of a Lie algebra $\frak{g}$ are labelled using the eigenvalues of
its generalized Casimir invariants \cite{Sh}. The number of
internal labels needed equals
\begin{equation}
i=\frac{1}{2}(\dim \frak{g}- \mathcal{N}(\frak{g})),
\end{equation}
as first observed by Racah \cite{Ra}. If we use some subalgebra
$\frak{h}$ to label the basis states of $\frak{g}$, we obtain
$\frac{1}{2}(\dim \frak{h}+\mathcal{N}(\frak{h})+l^{\prime}$
labels, where $l^{\prime}$ is the number of invariants of
$\frak{g}$ that depend only on variables of the subalgebra
$\frak{h}$ \cite{Sh}. In order to separate irreducible
representations of $\frak{g}$ uniquely, it is necessary to find
\begin{equation}
n=\frac{1}{2}\left(
\dim\frak{g}-\mathcal{N}(\frak{g})-\dim\frak{h}-\mathcal{N}(\frak{h})\right)+l^{\prime}
\label{ML}
\end{equation}
additional operators, which are usually called missing label
operators. The total number of available operators of this kind is
easily shown to be twice the number of needed labels, i.e.,
$m=2n$. For $n>1$, it remains the problem of determining a set of
$n$ mutually commuting operators. The analytical approach to the
missing label problem has the advantage of pointing out its close
relation to the problem of finding the invariants of the coadjoint
representation of a Lie algebra. Although in general the missing
label operators do not constitute invariants of the algebra or
subalgebra, they can actually be determined with the same Ansatz
\cite{Sh,Wi,C57}. Given the Lie algebra $\frak{g}$ with structure
tensor $\left\{ C_{ij}^{k}\right\} $ over a basis $\left\{
X_{1},..,X_{n}\right\} $, we realize the algebra  in the space
$C^{\infty }\left( \frak{g}^{\ast }\right) $ by means of the
differential operators defined by:
\begin{equation}
\widehat{X}_{i}=C_{ij}^{k}x_{k}\frac{\partial }{\partial x_{j}},
\label{Rep1}
\end{equation}
where $\left[ X_{i},X_{j}\right] =C_{ij}^{k}X_{k}$ \ $\left( 1\leq
i<j\leq n\right) $ and $\left\{ x_{1},..,x_{n}\right\}$ is a dual
basis of $\left\{X_{1},..,X_{n}\right\} $. The invariants of
$\frak{g}$ (in particular, the Casimir operators) are solutions of
the following system of partial differential equations:
\begin{equation}
\widehat{X}_{i}F=0,\quad 1\leq i\leq n.  \label{sys}
\end{equation}
Whenever we have a polynomial solution of (\ref{sys}), the
symmetrization map defined by
\begin{equation}
Sym(x_{i_{1}}^{a_{1}}..x_{i_{p}}^{a_{p}})=\frac{1}{p!}\sum_{\sigma\in
S_{p}}x_{\sigma(i_{1})}^{a_{1}}..x_{\sigma(i_{p})}^{a_{p}}
\end{equation}
allows to recover the Casimir operators in their usual form, i.e,
as elements in the centre of the enveloping algebra of $\frak{g}$.
A maximal set of functionally independent invariants is usually
called a fundamental basis. The number $\mathcal{N}(\frak{g})$ of
functionally independent solutions of (\ref{sys}) is obtained from
the classical criteria for differential equations, and is given
by:
\begin{equation}
\mathcal{N}(\frak{g}):=\dim \,\frak{g}- {\rm rank}\left(
C_{ij}^{k}x_{k}\right), \label{BB}
\end{equation}
where $A(\frak{g}):=\left(C_{ij}^{k}x_{k}\right)$ is the matrix
associated to the commutator table of $\frak{g}$ over the given
basis.  If we now consider an algebra-subalgebra chain $\frak{s}\supset\frak{s}%
^{\prime}$ determined by the embedding $f$, in order to compute
the missing label operators we have to consider the equations of
(\ref{sys}) corresponding to the generators of the subalgebra
$\frak{s}^{\prime}$. This system, as proven in \cite{Sh}, has
exactly $\mathcal{N}(f(\frak{s}^{\prime}))=\dim \frak{s}-\dim
\frak{s}^{\prime}-l^{\prime}$ solutions. Using formula (\ref{ML})
it follows further that this scalar can be expressed in terms of
the number of invariants of the algebra-subalgebra chain:
\begin{equation}
\mathcal{N}(f(\frak{s}^{\prime}))=m+\mathcal{N}(\frak{s})+\mathcal{N}(
\frak{s}^{\prime})-l^{\prime}. \label{ML2}
\end{equation}
This shows that the differential equations corresponding to the
subalgebra generators have exactly $n$ more solutions as needed to
solve the missing label problem. We remark that the scalar $m$
depends essentially on the embedding $f$.

\medskip

Since we are interested in combining the invariants with
contractions, we briefly recall the elementary notions that will
be used in the following. Let $\frak{g}$ be  a Lie algebra and
$\Phi_{t}\in End(\frak{g})$ a family of non-singular linear maps,
where $t\in [1,\infty)$.\footnote{Other authors use the parameter
range $(0,1]$, which is equivalent to this by simply changing the
parameter to $t^{\prime}=1/t$.} For any $X,Y\in\frak{g}$ we define
\begin{equation}
\left[X,Y\right]_{\Phi_{t}}:=\Phi_{t}^{-1}\left[\Phi_{t}(X),\Phi_{t}(Y)\right],
\end{equation}
which obviously represent the brackets of the Lie algebra over the
transformed basis. Now suppose that the limit
\begin{equation}
\left[X,Y\right]_{\infty}:=\lim_{t\rightarrow
\infty}\Phi_{t}^{-1}\left[\Phi_{t}(X),\Phi_{t}(Y)\right]
\label{Ko}
\end{equation}
exists for any $X,Y\in\frak{g}$. Then equation (\ref{Ko}) defines
a Lie algebra $\frak{g}^{\prime}$ called the contraction of
$\frak{g}$ (by $\Phi_{t}$), non-trivial if $\frak{g}$ and
$\frak{g}^{\prime}$ are non-isomorphic, and trivial otherwise
\cite{IW,We}. A contraction for which there exists some basis
$\left\{X_{1},..,X_{n}\right\}$ such that the contraction matrix
$A_{\Phi}$ is diagonal, that is, adopts the form
\begin{equation}
(A_{\Phi})_{ij}= \delta_{ij}t^{n_{j}},\quad n_{j}\in\mathbb{Z},
t>0, \label{IWK}
\end{equation}
is  called a generalized In\"on\"u-Wigner contraction \cite{We}.
This is the only type of contractions that we will need in this
work. It is known (see e.g. \cite{C23}) that for a contraction
$\frak{g}\rightsquigarrow \frak{g}^{\prime}$ of Lie algebras, the
following inequality must be satisfied
\begin{equation}
\mathcal{N}\left( \frak{g}\right)  \leq\mathcal{N}\left(
\frak{g}^{\prime }\right). \label{KB}
\end{equation}

The notion of contraction can also be formulated for invariant
functions \cite{We2}. The procedure is formally valid for
polynomial and non-polynomial invariants, but in this work we will
only consider Casimir operators. Suppose that the contraction is
of the type (\ref{IWK}). If
$F(X_{1},...,X_{n})=\alpha^{i_{1}...i_{p}}X_{i_{1}}...X_{i_{p}}$
is a Casimir operator of degree $p$, then the transformed
invariant takes the form
\begin{equation}
F(\Phi_{t}(X_{1}),..,\Phi_{t}(X_{n}))=t^{n_{i_{1}}+...+n_{i_{p}}}\alpha^{i_{1}...i_{p}}X_{i_{1}}...X_{i_{p}}.
\end{equation}
Now, defining
\begin{equation}
M=\max \left\{n_{i_{1}}+...+n_{i_{p}}\quad |\quad
\alpha^{i_{1}..i_{p}}\neq 0\right\},
\end{equation}
the limit
\begin{equation}
\fl F^{\prime}(X_{1},..,X_{n})=\lim_{t\rightarrow \infty}
t^{-M}F(\Phi_{t}(X_{1}),...,\Phi_{t}(X_{n}))=\sum_{n_{i_{1}}+...+n_{i_{p}}=M}
\alpha^{i_{1}...i_{p}}X_{i_{1}}...X_{i_{p}}
\end{equation}
gives a Casimir operator of degree $p$ of the contraction
$\frak{g}^{\prime}$. It should be remarked that, starting from an
adequate fundamental system of invariants
$\left\{C_{1},..,C_{p}\right\}$ of $\frak{g}$, it is always
possible to obtain a set of $p$ independent invariants of the
contraction. However, it is not ensured that these invariants are
of minimal degree in the contraction \cite{C62}.

\section{Embedding of Lie algebras and the associated contraction}

An embedding of a Lie algebra $\frak{s}^{\prime}$ into a Lie
algebra $\frak{s}$ is specified by an isomorphic mapping
$f:\frak{s}^{\prime}\longrightarrow \frak{s}$. A special type of
embeddings correspond to the so-called regular subalgebras, which
can be directly obtained from the Dynkin diagram of semisimple Lie
algebras \cite{Dy}. Each embedding determines an embedding index
$j_{f}$ and a  branching rule for irreducible representations of
$\frak{s}$, which depend essentially on the embedding. For simple
complex Lie algebras and maximal semisimple subalgebras, the
branching rules have been computed and tabulated up to rank eight
\cite{Mc}. In particular, for the reduction chain
$\frak{s}^{\prime}\hookrightarrow_{f}\frak{s}$, the adjoint
representation of $\frak{s}$ satisfies the following decomposition
\begin{equation}
{\rm ad} \frak{s}= {\rm ad} \frak{s}^{\prime}\oplus R,\label{D1}
\end{equation}
where $R$ is a (completely reducible) representation of
$\frak{s}^{\prime}$ determined by the embedding index
$j_{f}$.\footnote{The complete reducibility is actually ensured
only if the subalgebra $\frak{s}^{\prime}$ is semisimple.}

\medskip

In this paragraph we point out that any embedding of (semisimple)
Lie algebras $\frak{s}^{\prime}\subset \frak{s}$ naturally induces
a contraction of $\frak{s}$ onto an affine Lie algebra. To this
extent, consider a basis $\left\{
X_{1},..,X_{s},X_{s+1},..,X_{n}\right\} $ of $\frak{s}$ such that
$\left\{  X_{1},..,X_{s}\right\}  $ is a basis of
$\frak{s}^{\prime}$, and $\left\{ X_{s+1},..,X_{n}\right\}  $
spans the representation space of the induced $R$. Over this
basis, the structure tensor of $\frak{s}$ can be rewritten as
follows
\begin{eqnarray}
\left[  X_{i},X_{j}\right]& =\sum_{k=1}^{s}C_{ij}^{k}X_{k},\;1\leq
i,j,k\leq s,\\
\left[  X_{i},X_{j}\right]&
=\sum_{k=s+1}^{n}C_{ij}^{k}X_{k},\;1\leq i\leq
s,\;s+1\leq j,k \leq n,\\
\left[  X_{i},X_{j}\right] &=\sum_{k=1}^{s}C_{ij}^{k}X_{k}+\sum
_{l=s+1}^{n}C_{ij}^{l}X_{l},\;s+1\leq i,j\leq n.
\end{eqnarray}
For any $t\in\mathbb{R}$ we consider the non-singular linear transformations%
\begin{equation}
\Phi_{t}\left(  X_{i}\right)  =\left\{
\begin{array}
[c]{cc}%
X_{i}, & 1\leq i\leq s\\
\frac{1}{t}X_{i}, & s+1\leq i\leq n
\end{array}
\right.  .\label{TB}
\end{equation}
Expressing the brackets over the transformed basis $\left\{
X_{i}^{\prime }=\Phi_{t}\left(  X_{i}\right)  :\;1\leq i\leq
n\right\}  $ we obtain
\begin{eqnarray}
\left[  X_{i}^{\prime},X_{j}^{\prime}\right]    & =\sum_{k=1}^{s}C_{ij}%
^{k}X_{k}^{\prime},\;1\leq i,j,k\leq s,\\
\left[  X_{i}^{\prime},X_{j}^{\prime}\right]    & =\sum_{k=s+1}^{n}C_{ij}%
^{k}X_{k}^{\prime},\;1\leq i\leq s,\;s+1\leq j,k\leq n,\\
\left[  X_{i}^{\prime},X_{j}^{\prime}\right]    &
=\sum_{k=1}^{s}\frac
{1}{t^{2}}C_{ij}^{k}X_{k}^{\prime}+\sum_{l=s+1}^{n}\frac{1}{t}C_{ij}^{l}%
X_{l}^{\prime},\;s+1\leq i,j\leq n.
\end{eqnarray}
It follows at once that the subalgebra $\frak{s}^{\prime}$ remains
invariant, as well as the representation of $\frak{s}^{\prime}$
over its complementary in $\frak{s}$. These equations also show
that the limit
\[
\lim_{t\rightarrow\infty}\Phi_{t}^{-1}\left[  \Phi_{t}\left(
X\right) ,\Phi_{t}\left(  Y\right)  \right]
\]
exists for any pair of generators $X,Y\in\frak{s}$, we thus obtain
a non-trivial contraction\footnote{This is in fact a simple
In\"on\"u-Wigner contraction, following the notation of
\cite{We}.} of $\frak{s}$ denoted by $\frak{g}$ and with
non-vanishing brackets
\begin{eqnarray}
\left[  X_{i}^{\prime},X_{j}^{\prime}\right]    & =\sum_{k=1}^{s}C_{ij}%
^{k}X_{k}^{\prime},\;1\leq i,j,k\leq s,\\
\left[  X_{i}^{\prime},X_{j}^{\prime}\right]    & =\sum_{k=s+1}^{n}C_{ij}%
^{k}X_{k}^{\prime},\;1\leq i\leq s,\;s+1\leq j,k\leq n.
\end{eqnarray}
We observe that if $\frak{s}^{\prime}$ is semisimple, then it
coincides
with the Levi subalgebra of $\frak{g}$, and the Levi decomposition of this contraction equals%
\[
\frak{g}=\frak{s}^{\prime}\overrightarrow{\oplus}_{R}\left(
n-s\right) L_{1},
\]
where $(n-s)L_{1}$ denotes the Abelian algebra of dimension $n-s$.
This Lie algebra is affine, and by the contraction we know that
$\mathcal{N}(\frak{g})\geq \mathcal{N}(\frak{s})$. Applying the
analytical method, the invariants of $\frak{g}$ are obtained from
the solutions of the system:
\begin{eqnarray}
\widehat{X}_{i}F=C_{ij}^{k}x_{k}\frac{\partial F}{\partial
x_{j}}=0,\quad 1\leq i\leq s, \label{K1}\\
\widehat{X}_{s+i}F=C_{s+i,j}^{s+k}x_{s+k}\frac{\partial
F}{\partial x_{j}}=0,\quad 1\leq i,k\leq n-s,\; 1\leq j\leq
s.\label{K2}
\end{eqnarray}
Now equation (\ref{K1}) reproduces the subsystem of (\ref{sys})
corresponding to the generators of the embedded subalgebra
$\frak{s}^{\prime}$ that must be solved in order to find the
missing label operators for the reduction chain
$\frak{s}^{\prime}\subset \frak{s}$. This means in particular that
any invariant of the contraction $\frak{g}$ is a solution to that
system, thus can be taken as candidate for missing label operator,
whenever it is functionally independent from the invariants of
$\frak{s}$ and $\frak{s}^{\prime}$. As a consequence, we obtain
that $\mathcal{N}(f(\frak{s}^{\prime}))\geq
\mathcal{N}(\frak{g})$. Combining this inequality with formula
(\ref{ML2}), we conclude that
\begin{equation}
\mathcal{N}(f(\frak{s}))=m+\mathcal{N}(\frak{s})+\mathcal{N}(\frak{s}^{\prime})-l^{\prime}\geq
\mathcal{N}(\frak{g})\geq \mathcal{N}(\frak{s}).\label{ML3}
\end{equation}
The term $\mathcal{N}(f(\frak{s}))$ on the left hand side gives
the total number of available labelling operators, the invariants
of $\frak{s}$ and $\frak{s}^{\prime}$ comprised, as shown in
\cite{Sh}. Therefore, if the contraction $\frak{g}$ has enough
invariants, we can extract a set of $n$ commuting missing label
operators and solve the missing label problem completely. The most
important case in physical applications corresponds to reductions
chains of the type $\frak{s}\supset\frak{s}^{\prime}$, where
$\frak{s}$ is semisimple and $\frak{s}^{\prime}$ is a reductive
Lie algebra. Although the contraction method remains completely
valid for reductions involving non-reductive algebra-subalgebra
chains, in the following we will restrict ourselves to the case of
reductive subalgebras, for being the most representative case in
Physics.

\medskip

Suppose therefore that $\frak{s}$ is of rank $p$,
$\frak{s}^{\prime}$ is a
reductive subalgebra and let $\frak{g=s}^{\prime}\overrightarrow{\oplus}%
_{R}(\dim\frak{s}-\dim\frak{s}^{\prime})kL_{1}$ denote the contraction associated to the chain $\frak{s}%
\supset\frak{s}^{\prime}$. Let $\left\{  C_{1},..,C_{p}\right\}  $
be the Casimir operators of $\frak{s}$, and $\left\{
D_{1},..,D_{q}\right\}  $ the invariants of $\frak{s}^{\prime}$.
Contracting the invariants $C_{i}$ or some appropriate combination
of them, we can always obtain $p$ independent invariants of
$\frak{g}$. Completing if necessary to a maximal set of invariants
of $\frak{g}$, we obtain the fundamental system  $\left\{
C_{1}^{\prime},..,C_{p}^{\prime},..,C_{r}^{\prime}\right\}
\;$($r\geq p$). In order to solve the missing label problem using
the latter\ set of functions, the system $\mathcal{F}=\left\{
C_{1}^{\prime },..,C_{r}^{\prime}\right\}  $  must contain at
least $n$ functions that are independent on the Casimir invariants
of $\frak{s}$ and $\frak{s}^{\prime}$, i.e.,
\begin{equation}
{\rm rank}\,\mathcal{F}\; \left(  {\rm mod} \left\{  C_{1},..,C_{p}%
,D_{1},..D_{q}\right\}  \right)  \geq n.\label{U1}%
\end{equation}
By the construction, the set $\left\{
C_{1},..,C_{p},D_{1},..,D_{q-l^{\prime}}\right\}  $ is
functionally independent. Now the question arises whether adding
the invariants of $\frak{g}$ some dependence relations appear. In
general, and whenever no invariant is preserved by the
contraction, the functions $C_{i}$ and $C_{i}^{\prime}$ are
independent. In this case a dependence relation means that some
$C_{i}$ is a function of $C_{i}^{\prime}$ and the invariants of
$\frak{s}^{\prime}$. We observe that such a dependence relation
appears at least for the quadratic Casimir operator
$C_{1}$.\footnote{Is either $\frak{s}$ or $\frak{s}^{\prime}$ is
not reductive, this is not applicable, since existence of
quadratic operators is not ensured.} Indeed, writing $C_{1}$ over
the transformed basis (\ref{TB}) we obtain the following
decomposition of $C_{1}$ as polynomial in the contraction variable
$t$:
\[
C_{1}=F+t^{2}C_{1}^{\prime},
\]
where $F$ is a quadratic invariant of $\frak{s}^{\prime}$. This
decomposition follows from the well known fact that, over the
given basis, the quadratic Casimir operator of a reductive
subalgebra is always a summand of the quadratic Casimir operator
of $\frak{s}$.\footnote{For higher order invariants, dependence
relations could also appear, depending on the homogeneity degree
of the invariants of $\frak{s}$ with respect to the generators of
the subalgebra.} As a consequence, we obtain the upper bound
\begin{equation}
{\rm rank}\left\{  C_{1},..,C_{p},C_{1}^{\prime},..,C_{r}^{\prime},D_{1}%
,..,D_{q}\right\}  <\mathcal{N}\left(  \frak{g}\right)
+\mathcal{N}\left( \frak{s}\right)  +\mathcal{N}\left(
\frak{s}^{\prime}\right)  -l^{\prime
}.\label{U2}%
\end{equation}
Combining the lower and upper bounds (\ref{U1}) and (\ref{U2})
respectively, we obtain a necessary numerical condition on the
number of invariants of the contraction $\frak{g}$:
\begin{equation}
n<\mathcal{N}\left(  \frak{g}\right).  \label{U3}%
\end{equation}

These facts, put together, allow us to characterize when the
contraction $\frak{g}$ provides enough labelling operators to
solve the missing label problem for $\frak{s}\supset
\frak{s}^{\prime}$.

\begin{theorem}
A necessary and sufficient condition for solving the missing label
problem for the reduction $\frak{s}\supset\frak{s}^{\prime}$ by
means of the invariants of the associated contraction
$\frak{s\rightsquigarrow g=s}$ is that the affine Lie algebra
$\frak{g}$ satisfies the constraints

\begin{enumerate}
\item $\mathcal{N}\left(\frak{g}\right)  \geq n+1$,

\item  there are at least $n$ invariants of $\frak{g}$ that are
functionally independent from the invariants of $\frak{s}$ and
$\frak{s}^{\prime}$.
\end{enumerate}
\end{theorem}

The first condition, the easiest to evaluate, provides a numerical
criterion to decide whether the missing labels can be found by
means of the affine algebra $\frak{g}$. Unfortunately, there is no
general criterion to decide automatically whether and how many of
the contracted invariants are independent on the Casimir operators
of $\frak{s}$ and $\frak{s}^{\prime}$. We can however derive the
following sufficient condition.

\begin{corollary}
If the contraction $\frak{g}$ satisfies the numerical condition
$\mathcal{N}(\frak{g})\geq \left\{n+1,
\mathcal{N}(\frak{s})+\mathcal{N}(\frak{s}^{\prime})+1-l^{\prime}\right\}$,
then it solves the MLP.
\end{corollary}

\medskip

The use of the contraction naturally associated to an embedding
has further applications, which can be useful for a general study
of affine Lie algebras, in particular inhomogeneous algebras
\cite{Ch,Her,C49}. Let
$\frak{s}^{\prime}\hookrightarrow_{f_{1}}\frak{s}$ be an embedding
and
$\frak{s\rightsquigarrow g}=\frak{s}^{\prime}\overrightarrow{\oplus}_{R}%
kL_{1}$ the associated contraction. Since the subalgebra
$\frak{s}^{\prime}$ remains invariant by the contraction, we
naturally obtain the embedding
$f_{2}:\frak{s}^{\prime}\rightarrow\frak{g}$. If we now consider
the missing label problem for the latter
embedding,\footnote{Actually the mappings $f_{1}$ and $f_{2}$ are
the same, but we distinguish the target algebra by the indices.}
we immediately see that the system of PDEs to be solved is exactly
the same as for the embedding $f_{1}$. This means that the
solutions coincide, and, in particular, their number. This implies
that \ $\mathcal{N}\left( f_{1}\left( \frak{s}^{\prime}\right)
\right) =\mathcal{N}\left( f_{2}\left( \frak{s}^{\prime}\right)
\right) $. Recall that for each embedding the number of
independent solutions is given by
\begin{eqnarray}
\mathcal{N}\left(  f_{1}\left(  \frak{s}^{\prime}\right)  \right)
&
=\dim\frak{s}-\dim\frak{s}^{\prime}+l^{\prime},\nonumber \\
\mathcal{N}\left(  f_{2}\left(  \frak{s}^{\prime}\right)  \right)
& =\dim\frak{g}-\dim\frak{s}^{\prime}+l_{1}^{\prime}, \label{U4}
\end{eqnarray}
where $l_{1}^{\prime}$ denotes the number of common invariants of
$\frak{s}^{\prime}$ and $\frak{g}$. Since contractions preserve
the dimension, we conclude from formula (\ref{U4}) that
$l^{\prime}=l_{1}^{\prime}$, that is, the subalgebra
$\frak{s}^{\prime}$ has the same number of common invariants with
$\frak{s}$ than with the contraction $\frak{g}$. On the other
hand, using the reformulation (\ref{ML2})
\begin{eqnarray}
\mathcal{N}\left(  f_{1}\left(  \frak{s}^{\prime}\right)  \right)
& =m+\mathcal{N}\left(  \frak{s}\right)  +\mathcal{N}\left(
\frak{s}^{\prime
}\right)  -l^{\prime}\nonumber \\
\mathcal{N}\left(  f_{2}\left(  \frak{s}^{\prime}\right)  \right)
& =\widetilde{m}+\mathcal{N}\left(  \frak{g}\right)
+\mathcal{N}\left(
\frak{s}^{\prime}\right)  -l_{1}^{\prime}%
\end{eqnarray}
we deduce that
\begin{equation}
m-\widetilde{m}=\mathcal{N}\left(  \frak{g}\right)
-\mathcal{N}\left( s\right)  \geq0.\label{ML3}
\end{equation}
This result tells us that the number of available labelling
operators for the reduction chain
$\frak{s}\supset\frak{s}^{\prime}$ is always higher than that of
the chain $\frak{g}\supset\frak{s}^{\prime}$. Even more, the
inequality (\ref{ML3}) gives us a criterion to compute the number
of invariants of contractions in dependence of the available
missing label operators with respect to an invariant subalgebra.

\begin{proposition}
Let $\frak{s}\rightsquigarrow \frak{g}$ be a contraction such that
the subalgebra $\frak{s}^{\prime}$ is (maximal) invariant. Then
following equality holds:
\[
\mathcal{N}\left( \frak{g}\right)=\mathcal{N}\left(
s\right)+m-\widetilde{m},
\]
where $m$ and $\widetilde{m}$ is the number of available missing
label operators for the algebra subalgebra chain $\frak{s}\supset
\frak{s}^{\prime}$ and $\frak{g}\supset \frak{s}^{\prime}$,
respectively.
\end{proposition}

This result has useful applications, like the determination of the
number of invariants of some inhomogeneous Lie algebras. As a
particular case, we obtain the following upper bound
\begin{equation}
\mathcal{N}\left( \frak{g}\right)\leq \mathcal{N}\left(
s\right)+m. \label{ML5}
\end{equation}
This bound has an important interpretation, namely, that the
number of invariants of a contraction is, in some sense,
determined by the number of available missing label operators for
the missing label problem with respect to a maximal subalgebra of
$\frak{s}$ that remains invariant by the contraction. This fact
establishes a quite strong restriction to semidirect products of
semisimple and Abelian Lie algebras to appear as contractions of
semisimple Lie algebras \cite{C67}.

\section{The case $n=m=0$}

In the case of zero missing labels, the invariants of the
algebra-subalgebra chain provide a complete description of the
states. This situation is not uncommon for certain canonical
embeddings, such as the inclusions $\frak{so}(N)\subset
\frak{so}(N+1)$ of (pseudo)-orthogonal Lie algebras. Even if this
case is trivial, its interpretation in terms of the associated
contraction provides some interesting information concerning the
invariants of the contraction.

At first, if $m=0$, then by formula (\ref{ML3}) we have
$\mathcal{N}\left(  \frak{g}\right)=\mathcal{N}\left( s\right)$,
i.e., the contraction determined by the embedding
$\frak{s}\supset\frak{s}^{\prime}$ preserves the number of
invariants. It is worth to be observed that the converse does not
necessarily hold. Moreover, by formula (\ref{ML}), we have
\begin{equation}
0=m=\dim \frak{s}-\dim
\frak{s}^{\prime}-\mathcal{N}(\frak{s})-\mathcal{N}(\frak{s}^{\prime})+2l^{\prime}.
\end{equation}
In absence of additional internal labels, the system
$\widehat{X}_{i}F=0$ for the generators of $\frak{s}^{\prime}$ has
exactly
\begin{equation}
\mathcal{N}(f(\frak{s}))=\mathcal{N}(\frak{s})+\mathcal{N}(\frak{s}^{\prime})-l^{\prime}
\end{equation}
solutions. Since any invariant of the contraction $\frak{g}=\frak{s}^{\prime}\overrightarrow{\oplus}_{R}%
(\dim\frak{s}-\dim\frak{s}^{\prime})L_{1}$ is a special solution
of this system, the latter equation tells that any invariant of
$\frak{g}$ is functionally dependent on the invariants of
$\frak{s}$ and the subalgebra $\frak{s}^{\prime}$. That is, the
Casimir invariants of the algebra-subalgebra chain completely
determine the invariants of the contraction.\footnote{Of course,
if $\mathcal{N}(\frak{s}^{\prime})=0$, this assertion fails, but
for reductive subalgebras this situation is excluded.} Expressed
in another way, in this situation, polynomial functions of the
invariants of $\frak{s}$ and the contraction $\frak{g}$ allow to
recover naturally the invariants of the subalgebra.

\medskip

These observations provide a new (and very short) proof of the
fact that the number of invariants for inhomogeneous
pseudo-orthogonal Lie algebras is given by
\begin{equation}
\mathcal{N}(I\frak{so}(p,q))=\left[\frac{p+q+1}{2}\right].
\end{equation}
In fact, it is straightforward to verify that $n=0$, and since
$I\frak{so}(p,q)$ is a contraction of $\frak{so}(p+1,q)$, the
result follows at once from formula (\ref{ML5}). Moreover, the
invariants of $I\frak{so}(p,q))$ can be obtained from the
invariants of $\frak{so}(p+1,q)$ and $\frak{so}(p,q)$. This
explains in some manner why the classical Gel'fand method applies
so well to inhomogeneous algebras of this kind \cite{C49}.

\medskip

As example, consider the embedding $\frak{so}\left(  3,1\right)
\hookrightarrow\frak{so}\left(  4,1\right)  $ of the Lorentz
algebra into the Anti De Sitter algebra $\frak{so}\left(
4,1\right) $. Using the   kinematical  basis $\left\{
J_{\alpha},P_{\alpha},K_{\alpha},H\right\} _{1\leq\alpha\leq3}$,
where $J_{\alpha}$ are spatial rotations, $P_{\alpha}$ spatial
translations, $K_{\alpha}$ the boosts and $H$ the time
translation, the non-trivial brackets of $\frak{so}\left(
4,1\right)  $ are
\begin{equation}%
\fl
\begin{array}
[c]{llll}%
\left[  J_{\alpha},J_{\beta}\right]  =\varepsilon^{\alpha\beta\gamma}%
J_{\gamma}, & \left[  J_{\alpha},P_{\beta}\right]
=\varepsilon^{\alpha \beta\gamma}P_{\gamma}, & \left[
J_{\alpha},K_{\beta}\right]  =\varepsilon
^{\alpha\beta\gamma}K_{\gamma}, & \left[  H,P_{\alpha}\right]
=\varepsilon
^{\alpha\beta\gamma}K_{\alpha},\\
\left[  H,K_{\alpha}\right]
=\varepsilon^{\alpha\beta\gamma}P_{\alpha}, &
\left[  P_{\alpha},P_{\beta}\right]  =\varepsilon^{\alpha\beta\gamma}%
J_{\gamma}, & \left[  K_{\alpha},K_{\beta}\right]
=-\varepsilon^{\alpha \beta\gamma}J_{\gamma}, & \left[
P_{\alpha},K_{\alpha}\right]  =H.
\end{array}
\end{equation}
It follows at once that $\frak{so}\left(  3,1\right)  $ is
generated by the rotations and boosts. In this case there are no
missing labels, thus $n=m=0$.
The corresponding contraction defined by the linear maps%
\[
J_{\alpha}^{\prime}=J_{\alpha},P_{\alpha}^{\prime}=\frac{1}{t}P_{\alpha
},K_{\alpha}^{\prime}=K_{\alpha},H^{\prime}=\frac{1}{t}H
\]
leads to the Poincar\'{e} algebra $I\frak{so}\left(  3,1\right)
$. Over this
basis, the Casimir operators of $\frak{so}\left(  4,1\right)  $ are%
\begin{eqnarray*}
\fl
C_{2}   =j_{\alpha}j^{\alpha}+p_{\alpha}p^{\alpha}-k_{\alpha}k^{\alpha}%
-h^{2}\\
\fl C_{4}   =j_{\alpha}j^{\alpha}h^{2}+\left(
p_{\alpha}p^{\alpha}\right) \left(  k_{\alpha}k^{\alpha}\right)
-\left(  p_{\alpha}k^{\alpha}\right) ^{2}+\left(
p_{\alpha}j^{\alpha}\right)  ^{2}- \left(
j_{\alpha}k^{\alpha}\right) ^{2}-2\varepsilon^{\alpha\beta\gamma
}j_{\alpha}p_{\beta}k_{\gamma}h.
\end{eqnarray*}
Contraction of these invariants give the Casimir operators of the
Poincar\'{e}
algebra%
\begin{eqnarray*}
C_{2}^{\prime}  & =p_{\alpha}p^{\alpha}-h^{2}\\
C_{4}^{\prime}  & =j_{\alpha}j^{\alpha}h^{2}+\left(
p_{\alpha}p^{\alpha }\right)  \left(  k_{\alpha}k^{\alpha}\right)
-\left(  p_{\alpha}k^{\alpha }\right)  ^{2}+\left(
p_{\alpha}j^{\alpha}\right)  ^{2}-
2\varepsilon^{\alpha\beta\gamma}j_{\alpha}p_{\beta}k_{\gamma}h.
\end{eqnarray*}
Now observe that
$C_{21}=p_{\alpha}p^{\alpha}-k_{\alpha}k^{\alpha}$ and
$C_{22}=j_{a}k^{\alpha}$ are the Casimir operators of the
$\frak{so}\left( 3,1\right)  $ subalgebra. It follows that
\[
C_{2}=C_{2}^{\prime}+C_{21},\;C_{4}=C_{4}^{\prime}-C_{22}^{2},
\]
i.e., the mass squared and spin squared operators of the
Poincar\'{e} algebra are obtainable as a difference of the Casimir
operators of the Lorentz and De Sitter Lie algebras, and therefore
the information they provide is already contained in the reduction
chain.

\section{The case $n=1, m=2$}

In the case of one missing label operator, any solution of the
contraction $\frak{g}$ that is independent of the invariants of
the algebra-subalgebra chain can be used. No commutation problems
arise at this step. Formula (\ref{ML5}) establishes the maximal
possible number for the invariants of $\frak{g}$:
\[
\mathcal{N}\left( \frak{g}\right)\leq \mathcal{N}\left(
s\right)+2.
\]
For the case of semisimple Lie algebra $\frak{s}$ and maximal
reductive subalgebra $\frak{s}^{\prime}$, there are eight cases
with one missing label \cite{Sh,Lo}. Most of these chains have
been solved finding finite integrity bases, that is, a set of
elementary subgroup scalar such that any other can be expressed by
a polynomial in them. All eight cases can also be solved applying
the contraction method. In order to illustrate how the contraction
method works, we consider two representative cases, and resume the
results for the remaining cases in Table 1.

\subsection{The $\frak{su}\left(  3\right)  \supset\frak{so}\left(  3\right)
$ reduction}

This reduction chain, first considered in atomic physics by
Elliott, is probably the best known and best studied case
concerning the missing label problem. A complete set of commuting
operators and their eigenvalues for different irreducible
representations of $\frak{su}(3)$ were first determined in
\cite{Za}.\newline The $\frak{so}\left(  3\right)  $ subalgebra is
naturally identified with the three orbital angular momentum
operators, while the remaining five generators transform under
rotations like the elements of a second rank tensor \cite{El,Ro2}.
Here we consider a basis $\left\{  L_{i},T_{jk}\right\}  $
formed by rotations $L_{i}$ and the operators $T_{ik}$ and commutation relations%
\begin{equation*}
\begin{tabular}
[c]{ll}%
$\left[  L_{j},L_{k}\right]  =i\varepsilon_{jkl}L_{l},$ & $\left[
L_{j},T_{kl}\right]  =i\varepsilon_{jkm}T_{lm}+i\varepsilon_{jlm}T_{km},$\\
\multicolumn{2}{l}{$\left[  T_{jk},T_{lm}\right]
=\frac{i}{4}\left\{
\delta_{j}^{l}\varepsilon_{kmn}+\delta_{j}^{m}\varepsilon_{k\ln}+\delta
_{k}^{l}\varepsilon_{jmn}+\delta_{k}^{m}\varepsilon_{j\ln}\right\}  L_{n},$}%
\end{tabular}
\end{equation*}
where $T_{33}+\left(  T_{11}+T_{22}\right)  =0$. The symmetrized
Casimir operators, following the notation of \cite{Za}, are given
by $C^{\left(
2\right)  }=L_{i}L_{i}+2T_{ik}T_{ik}$, $\ C^{\left(  3\right)  }=L_{i}%
T_{ik}L_{k}-\frac{4}{3}T_{ik}T_{kl}T_{li}$ and $C^{\left(  2,0\right)  }%
=L_{i}L_{i}$. The contraction $\frak{g}$ associated to this
reduction has Levi
decomposition $\frak{g=so}\left(  3\right)  \overrightarrow{\oplus}_{R_{5}%
^{I}}5L_{1}$, where $R_{5}^{I}$ denotes the five dimensional
irreducible representation of $\frak{so}\left(  3\right)  $. This
 is equivalent to the rotor algebra $[\mathbb{R}^{5}]SO(3)$
 studied in \cite{Ro}. It is straightforward to verify that $\mathcal{N}\left(
\frak{g}\right) =2$. Therefore, a basis of invariants of
$\frak{g}$ can be obtained by contraction of $C^{\left( 2\right)
}$ and $C^{\left( 3\right)  }$. Specifically, we get the
(unsymmetrized) Casimir invariants%
\begin{eqnarray*}
C_{2}  =2t_{ik}t^{ik},\\
C_{3}  =t_{ik}t^{kl}t_{li}.
\end{eqnarray*}
As already observed, $C_{2}$ is functionally dependent on
$C^{\left( 2\right) }$ and $C^{\left(  2,0\right)  }$, therefore
of no use for the MLP. The independence of $\left\{  C^{\left(
2\right) },C^{\left( 3\right)  },C^{\left(
2,0\right)  },C_{3}\right\}  $ follows from the Jacobian%
\[
\frac{\partial\left\{  C^{\left(  2\right)  },C^{\left(  3\right)
},C^{\left(  2,0\right)  },C_{3}\right\}  }{\partial\left\{  l_{2}%
,l_{3},t_{11},t_{12}\right\}  }\neq0.
\]
The invariant $C_{3}$ is therefore sufficient to solve the missing
label problem. In fact, we can recover the missing label operator
$X^{\left( 3\right)  }$ from \cite{Za} by simply considering the
linear combination
\[
X^{\left(  3\right)  }=C^{\left(  3\right)  }+\frac{4}{3}\left\{
C_{3}\right\}  _{symmetrized}.
\]
This operator is equivalent to the third order operator obtained
by Bargmann and Moshinsky in \cite{Ba}, and also to the operator
determined in \cite{Ro} using the K-matrix approach. It is
observed that the fourth order operator $X^{\left( 4\right)
}=L_{i}T_{ij}T_{jk}L_{k}$ cannot be obtained from the invariants
of $\frak{su}\left( 3\right) ,\frak{so}\left( 3\right)  $ and the
contraction $\frak{g}$. This is essentially due to the fact that
the fundamental Casimir operators of $\frak{su}(3)$ have degree
two and three.

\subsection{The seniority model}

The reduction $\frak{so}\left(  5\right)  \supset\frak{su}\left(
2\right) \times\frak{u}\left(  1\right)  $ has been used in the
treatment of the paring force between particles in the same
nuclear shell, and is usually referred to as the seniority model
\cite{He}.

In order to analyze this chain, we use the same basis $\left\{
U_{\pm},U_{3},V_{3},V_{\pm},S_{\pm },T_{\pm}\right\}  $ of
\cite{Sh3}. The $\frak{su}\left(  2\right) \times\frak{u}\left(
1\right)  $ subalgebra is generated by the operators $\left\{
U_{\pm},U_{3},V_{3}\right\}  $. The nonzero brackets are given by
\begin{equation*}
\fl
\begin{tabular}
[c]{llll}%
$\left[  U_{\pm},U_{3}\right]  =\mp U_{\pm},$ & $\left[
U_{+},U_{-}\right] =2U_{3},$ & $\left[  U_{\pm},V_{\pm}\right]
=\mp2S_{\pm},$ & $\left[  U_{\pm
},V_{\mp}\right]  =\mp2T_{\pm},$\\
$\left[  U_{\pm},S_{\mp}\right]  =\pm V_{\mp},$ & $\left[
U_{\pm},T_{\mp }\right]  =\pm V_{\mp},$ & $\left[
U_{3},S_{\pm}\right]  =\pm S_{\pm},$ &
$\left[  U_{3},T_{\pm}\right]  =\pm T_{\pm},$\\
$\left[  V_{3},S_{\pm}\right]  =\pm S_{\pm},$ & $\left[
V_{3},T_{\pm}\right] =\mp T_{\pm},$ & $\left[  V_{+},V_{-}\right]
=2V_{3},$ & $\left[  V_{\pm
},V_{3}\right]  =\mp V_{\pm},$\\
$\left[  V_{\pm},S_{\mp}\right]  =\mp U_{\mp},$ & $\left[
V_{\pm},T_{\pm }\right]  =\pm U_{\pm},$ & $\left[
S_{+},S_{-}\right]  =U_{3}+V_{3},$ &
$\left[  T_{+},T_{-}\right]  =U_{3}-V_{3}.$%
\end{tabular}
\end{equation*}
Over this basis, the (unsymmetrized) Casimir operators of
$\frak{so}\left( 5\right)  $ can be chosen as
\begin{eqnarray*}
\fl C_{2} =u_{+}u_{-}+u_{3}^{2}+v_{3}^{2}+v_{+}v_{-}+2\left(  s_{+}s_{-}%
+t_{+}t_{-}\right)  ,\\
\fl C_{4}   =\left(  u_{+}u_{-}+u_{3}^{2}\right)
v_{3}^{2}+u_{+}u_{-}\left(
s_{+}s_{-}+t_{+}t_{-}\right)+u_{+}^{2}s_{-}t_{-}%
+u_{-}^{2}s_{+}t_{+}+2u_{3}v_{3}\left(  s_{+}s_{-}-t_{+}t_{-}\right)  \\
\fl +\left(  \left(  t_{-}v_{-}-s_{-}v_{+}\right)  u_{+}+\left(  t_{+}v_{+}%
-s_{+}v_{-}\right)  u_{-}\right)  v_{3}+\left(  \left(  t_{+}v_{+}+s_{+}%
v_{-}\right)  u_{-}+\left(  s_{-}v_{+}+t_{-}v_{-}\right)
u_{+}\right)u_{3}\\
\fl  +v_{+}v_{-}s_{+}s_{-}+u_{3}^{2}v_{+}v_{-}+ \left(  s_{+}s_{-}-t_{+}
t_{-}\right)  ^{2}-v_{+}^{2}s_{-}t_{+}-v_{-}%
^{2}s_{+}t_{-}+v_{+}v_{-}t_{+}t_{-},
\end{eqnarray*}
while those of the subalgebra are given by
$C_{21}=u_{+}u_{-}+u_{3}^{2},\;C_{22}=v_{3}$. The associated
contraction $\frak{g}$ is easily seen to have exactly two
invariants, which can be obtained from those of $\frak{so}\left(
5\right)  $
by the contraction method:%
\begin{eqnarray*}
\fl C_{2}^{\prime}  & =v_{+}v_{-}+2\left(  s_{+}s_{-}+t_{+}t_{-}\right)  ,\\
\fl C_{4}^{\prime}  & =v_{+}v_{-}s_{+}s_{-}+\left(
s_{+}s_{-}-t_{+}t_{-}\right)
^{2}-v_{+}^{2}s_{-}t_{+}-v_{-}^{2}s_{+}t_{-}+v_{+}v_{-}t_{+}t_{-}.
\end{eqnarray*}
As expected, we have $C_{2}=C_{2}^{\prime}+C_{21}+C_{22}^{2}$,
thus at most $C_{4}^{\prime}$ is independent on the invariants of
$\frak{so}\left( 5\right)  \,$\ and $\frak{su}\left(  2\right)
\times\frak{u}\left(  1\right) $. A short computation shows that
\[
{\rm rank }\left\{
C_{2},C_{4},C_{21},C_{22},C_{4}^{\prime}\right\} =5,
\]
showing that the missing label problem can be solved using the
contraction $\frak{g}$. Now, after some manipulation we can arrive
at the expression
$\Omega_{4}=C_{4}-C_{4}^{\prime}-C_{21}C_{22}^{2}$ explicitly
given by
\begin{eqnarray*}
\fl \Omega_{4}  =u_{+}u_{-}\left(  s_{+}s_{-}+t_{+}t_{-}\right)  +u_{3}^{2}%
v_{+}v_{-}+u_{+}^{2}s_{-}t_{-}+u_{-}^{2}s_{+}t_{+}+2u_{3}v_{3}\left(
s_{+}s_{-}-t_{+}t_{-}\right) \\
\fl +\left(  \left(  t_{-}v_{-}-s_{-}v_{+}\right)  u_{+}+\left(  t_{+}v_{+}%
-s_{+}v_{-}\right)  u_{-}\right)  v_{3}+\left(  \left(  t_{+}v_{+}+s_{+}%
v_{-}\right)  u_{-}+\left(  s_{-}v_{+}+t_{-}v_{-}\right)
u_{+}\right)  u_{3}.
\end{eqnarray*}

This operator is obviously independent on the invariants of the
orthogonal algebra and the subalgebra, and can therefore be taken
as the missing operator. It can be verified that $\Omega_{4}$,
after symmetrization, coincides with the fourth order operator
$UVL^{2}$ found in \cite{Sh}. The remaining third order operator
cannot be obtained using the contraction $\frak{g}$. In this case,
this is a consequence of the non-existence of cubic Casimir
operators for the orthogonal algebra $\frak{so}(5)$.

\begin{table}
\caption{\label{Tab2} Comparison of missing labels of \cite{Sh}
and those obtained by contraction. }
\begin{indented}
\item[]\begin{tabular}[c]{@{}l|ccccc} $\frak{s\supset s}^{\prime}$
& $\mathcal{N}\left(  \frak{g}\right)  $ & $\mathcal{N}\left(
f\left(  \frak{s}^{\prime}\right)  \right)  $ & rank $\mathcal{F}$
& Order of $\Phi$ & Operator of \cite{Sh}\\\hline $\frak{su}\left(
3\right)  \supset\frak{so}\left(  3\right)  $ & $2$ & $5$ &
$4$ & $3$ & $X^{(3)}$$^{\rm a}$\\
$\frak{so}\left(  5\right)  \supset\frak{su}\left(  2\right)
\times
\frak{u}\left(  1\right)  $ & $2$ & $6$ & $5$ & $4$ & $UVL^{2}$\\
$G_{2}\supset\frak{su}\left(  3\right)  $ & $2$ & $5$ & $4$ & $6$ & $U^{3}V^{3}%
$\\
$\frak{sp}\left(  6\right)  \supset\frak{sp}\left(  4\right)
\times
\frak{su}\left(  2\right)  $ & $3$ & $8$ & $7$ & $6$ & $Q^{3}T^{2}L$\\
$\frak{so}\left(  7\right)  \supset G_{2}$ & $3$ & $7$ & $6$ & $6$
&
$T^{4}S^{2}$\\
$\frak{su}\left(  4\right)  \supset\left[  \frak{su}\left(
2\right)  \right]
^{2}\times\frak{u}\left(  1\right)  $ & $3$ & $7$ & $6$ & $4$ & $UVST$\\
$\frak{su}\left(  3\right)  \times\frak{su}\left(  3\right)
\supset
\frak{su}\left(  3\right)  $ & $2$ & $8$ & $7$ & $3$ & $UV^{2}$\\
$\left[\frak{su}\left(  2\right)\right]^{3}\supset\frak{su}\left(
2\right) $ & $3$ & $6$ & $5$ & $2$ & $-^{\rm b}$\\\hline\br
\end{tabular}
\item[] $^{\rm a}$ The notation for the operator corresponds to
that used in \cite{Za}. \item[]$^{\rm b}$ This case, omitted in
\cite{Sh}, was first considered in \cite{Lo}.

\end{indented}
\end{table}

\section{The case $n=2, m=4$}

The case with two missing labels is notably more complicated,
because in addition to determine two missing label operators,
these must commute. Although a considerable number of cases has
been studied, only for a few the most general form of missing
label operators has been analyzed in detail, such as the Wigner
supermultiplet $\frak{su}(4)\supset
\frak{su}(2)\times\frak{su}(2)$ \cite{Que,Pa} or the chain
$\frak{so}(5)\supset \frak{su}(2)$ used for the classification of
nuclear surfon states \cite{He}.

\subsection{ The supermultiplet model}

This model, used by Wigner to describe light nuclei, has been
considered in detail by various authors, usually by means of
enveloping algebras \cite{Que,Pa,Mo}. It has been shown that the
set of available operators is partitioned into two separate sets,
the Moshinky-Nagel operators ${\Omega,\Phi}$ and two other
operators ${O_{1},O_{2}}$, first found in \cite{Que} and later
evaluated numerically in \cite{Pa}. We start from the same basis
$\left\{ S_{i},T_{j},Q_{\alpha\beta}\right\}  $ used in \cite{Mo},
where $1\leq i,j,\alpha,\beta\leq3$. The non-vanishing
brackets of $\frak{su}\left(  4\right)  $ are%
\begin{eqnarray}
\fl \left[  S_{i},S_{j}\right]  =i\varepsilon_{ijk}S_{k},\; \left[  T_{i}%
,T_{j}\right]  =i\varepsilon_{ijk}T_{k},\; \left[
S_{i},Q_{j\alpha}\right] =i\varepsilon_{ijk}Q_{k\alpha},\; \left[
T_{\alpha},Q_{i\beta}\right]
=i\varepsilon_{\alpha\beta\gamma}Q_{i\gamma},\nonumber\\
\fl \left[  Q_{i\alpha},Q_{j\beta}\right]  =\frac{i}%
{4}\left\{  \delta_{\alpha\beta}\varepsilon_{ijk}S_{k}+\delta_{ij}%
\varepsilon_{\alpha\beta\gamma}T_{\gamma}\right\} ,
\end{eqnarray}
where $\varepsilon_{ijk}$ is the completely antisymmetric tensor.
The $\frak{su}\left(  2\right)  \times\frak{su}\left(  2\right)
$-subalgebra is generated by the operators $\left\{
S_{i},T_{j}\right\}  $.\ It follows easily from the  brackets that
the generators of $\frak{su}\left(  4\right)  $ decompose as the
following $\frak{su}\left(  2\right)  \times\frak{su}\left(
2\right)  $-representation
\begin{equation}
R=\left(  D_{1}\otimes D_{0}\right)  \oplus\left(  D_{0}\otimes
D_{1}\right)
\oplus\left(  D_{1}\otimes D_{1}\right)  ,\label{RD}%
\end{equation}
where $D_{1}$ denotes the adjoint representation of $\frak{su}(2)$
and $D_{0}$ the trivial representation. The two missing label
operators are therefore determined by the system of differential
equations
\begin{equation}
\fl \widehat{S}_{i}F   =\epsilon_{ijk}s_{k}\frac{\partial F}{\partial s_{j}%
}+\epsilon_{ijk}q_{kl}\frac{\partial F}{\partial q_{kl}}=0,\quad
\widehat{T}_{\alpha}F
=\epsilon_{\alpha\beta\gamma}t_{\gamma}\frac{\partial F}{\partial
t_{\beta}}+\epsilon_{\beta\gamma\mu}q_{\alpha\mu}\frac{\partial
F}{\partial q_{\beta\mu}}=0,\quad i=1,2,3\label{S2}%
\end{equation}
corresponding to the generators of the subalgebra. From the nine
independent solutions, five of them correspond to invariants of
$\frak{su}\left(
4\right)  $ and the subalgebra. The Casimir operators can be taken as$.$%
\begin{eqnarray}
\fl
C_{2}   =s_{\alpha}s^{\alpha}+t_{\beta}t^{\beta}+4q_{\alpha\beta}%
q^{\alpha\beta},\\
\fl C_{3}
=s_{\alpha}t_{\beta}q^{\alpha\beta}-4\varepsilon^{ijk}\varepsilon
^{\alpha\beta\gamma}q_{i\alpha}q_{j\beta}q_{k\gamma},\\
\fl C_{4}   =16\left\{  \varepsilon_{\alpha\beta\gamma}^{2}(q_{\alpha\beta}%
^{2}\left(  q_{\alpha\gamma}^{2}+q_{\gamma\beta}^{2}\right)
+2q_{\alpha \alpha}^{2}\left(
q_{\alpha\gamma}^{2}+q_{\beta\alpha}^{2}\right)
-2q_{\alpha\alpha}q_{\alpha\beta}q_{\gamma\alpha}q_{\gamma\beta}+\left.
3q_{\alpha\beta}^{2}\left( q_{\gamma\alpha}^{2}+q_{\gamma\gamma
}^{2}\right)  \right)\right.  \nonumber\\
\fl \left. +\sum_{a<\beta}\left(  3\left(  q_{\alpha\alpha}^{2}%
q_{\beta\beta}^{2}+q_{\alpha\beta}^{2}q_{\beta\alpha}^{2}\right)
-2q_{\alpha\alpha}q_{\beta\beta}q_{\alpha\beta}q_{\beta\alpha}\right)
+q_{\alpha\beta}^{4}\right\} +\left( s_{\alpha}s^{\alpha}\right)
^{2}+\left(
t_{\beta}t^{\beta}\right)  ^{2}+3s_{\alpha}s^{\alpha}t_{\beta}t^{\beta}%
\nonumber\\
\fl +2^{3}q_{\alpha\beta}^{2}\left(
s_{\alpha}s^{\alpha}+t_{\beta}t^{\beta }\right)  +4\left\{
t_{\alpha}t_{\beta}q_{\gamma\alpha}q_{\gamma\beta}+s_{\alpha
}s_{\beta}q_{\alpha\gamma}q_{\beta\gamma}-\varepsilon_{\alpha\beta\gamma
}\varepsilon_{\mu\nu\rho}s_{\mu}t_{\alpha}q_{\nu\beta}q_{\rho\gamma}\right\}
\end{eqnarray}
for $\frak{su}\left(  4\right)$, and
$C_{21}=s_{\alpha}s^{\alpha},\quad C_{22}=t_{\beta}t^{\beta}$ for
the subalgebra. In this case, the contraction
$\frak{g}=(\frak{su}(2)\times\frak{su}(2))\overrightarrow{\oplus}%
_{D_{1}\otimes D_{1}}9L_{1}$ associated to the embedding has the
following
non-trivial brackets%
\begin{equation}
\fl \left[  S_{i},S_{j}\right]  =i\varepsilon_{ijk}S_{k},\; \left[  T_{i}%
,T_{j}\right]  =i\varepsilon_{ijk}T_{k},\; \left[
S_{i},Q_{j\alpha}\right] =i\varepsilon_{ijk}Q_{k\alpha},\; \left[
T_{\alpha},Q_{i\beta}\right]
=i\varepsilon_{\alpha\beta\gamma}Q_{i\gamma}.
\end{equation}
Using formula (\ref{BB}) we easily get $\mathcal{N}\left(
\frak{g}\right) =3$. Contracting the invariants we obtain three
independent invariants of $\frak{g}$, given respectively by
\begin{eqnarray}
\fl C_{2}^{\prime} =4q_{\alpha\beta}q^{\alpha\beta},\\
\fl C_{3}^{\prime}
=-4\varepsilon^{ijk}\varepsilon^{\alpha\beta\gamma
}q_{i\alpha}q_{j\beta}q_{k\gamma},\\
\fl C_{4}^{\prime}   =16\left\{  \varepsilon_{\alpha\beta\gamma}^{2}%
(q_{\alpha\beta}^{2}\left(
q_{\alpha\gamma}^{2}+q_{\gamma\beta}^{2}\right)
+2q_{\alpha\alpha}^{2}\left(
q_{\alpha\gamma}^{2}+q_{\beta\alpha}^{2}\right)
-2q_{\alpha\alpha}q_{\alpha\beta}q_{\gamma\alpha}q_{\gamma\beta}%
+3q_{\alpha\beta}^{2}\left( q_{\gamma\alpha}^{2}+q_{\gamma\gamma
}^{2}\right)  )\right.  \nonumber\\
\fl \left. +q_{\alpha\beta}^{4}+\sum_{a<\beta}\left(  3\left(  q_{\alpha\alpha}^{2}%
q_{\beta\beta}^{2}+q_{\alpha\beta}^{2}q_{\beta\alpha}^{2}\right)
-2q_{\alpha\alpha}q_{\beta\beta}q_{\alpha\beta}q_{\beta\alpha}\right)
\right\}  .
\end{eqnarray}
As observed, the quadratic Casimir operator of $\frak{g}$
satisfies the condition $C_{2}-C_{2}^{\prime}=C_{21}+C_{22}$, and
is therefore dependent. To
prove that $\mathcal{F}=\left\{  C_{2},C_{3},C_{4},C_{21},C_{22},C_{3}%
^{\prime},C_{4}^{\prime}\right\}  $ is a functionally independent
set, we consider
the Jacobian with respect to the variables $\left\{  s_{2},s_{3},t_{1}%
,t_{2},q_{11},q_{12},q_{23}\right\}  $ :
\begin{equation}
\frac{\partial(C_{21},C_{2},C_{3},C_{4},C_{2}^{\prime},C_{3}^{\prime}%
,C_{4}^{\prime})}{\partial(s_{2},s_{3},t_{1},t_{2},q_{11},q_{12},q_{23})}%
\neq0.
\end{equation}
Actually, this is a maximal set of independent functions among the
invariants of the intervening Lie algebras $\frak{su}\left(
4\right)  ,\frak{su}\left( 2\right)  \times\frak{su}\left(
2\right)  $ and $\frak{g}$. This means that the contraction method
provides at most two of the four available operators. If we take
the difference of the cubic invariants of $\frak{su}(4)$ and
$\frak{g}$, we recover exactly the cubic operator $\Omega$ of
Moshinsky and Nagel\ \cite{Mo}:
\begin{equation}
C_{3}-C_{3}^{\prime}=\Omega=s_{\alpha}t_{\beta}q^{\alpha\beta}.
\end{equation}
As known, the operator  $\Omega$ only commutes with the fourth
order operator $\Phi$ defined by
\begin{equation}
\Phi=S_{i}S_{j}Q_{i\alpha}Q_{j\alpha}+Q_{i\alpha}Q_{i\beta}T_{\alpha}T_{\beta
}-\epsilon_{ijk}\epsilon_{\alpha\beta\gamma}S_{i}T_{\alpha}Q_{j\beta
}Q_{k\gamma}.\label{Op}%
\end{equation}
With some more effort we can express $\Phi$ with the help of the
preceding functions of $\mathcal{F}$, obtaining
\begin{equation}
\Phi=\frac{1}{4}\left\{  C_{4}-C_{4}^{\prime}+C_{21}^{2}-C_{2}^{2}%
+C_{2}^{\prime2}-C_{21}\left(  C_{2}^{\prime}-C_{2}\right)
\right\}
.\label{Mos1}%
\end{equation}
This means that the commuting $\Omega-\Phi$ operators of
Moshinky-Nagel are completely determined by the contraction
associated to the embedding of spin-isospin subalgebra in
$\frak{su}\left(  4\right)  $, while the other pair of commuting
operators, being summands of $\Phi$, cannot be obtained by this
method.

\subsection{The nuclear surfon model}

The reduction chain $\frak{so}(5)\supset \frak{su}(2)$ has been
analyzed in \cite{Me}, where two commuting missing label operators
of degrees four and six were found. The authors looked for the
simplest possible operators solving the labelling problem. We
reconsider the problem with the contraction method. As in
\cite{Me}, we choose the basis of $\frak{so}(5)$ to consist of
generators $\left\{L_{0},L_{1},L_{-1}\right\}$ with brackets
$\left[L_{0},L_{\pm 1}\right]=\pm L_{\pm 1},\;
\left[L_{1},L_{-1}\right]=2L_{0}$ together with an irreducible
tensor representation $Q_{\mu}\; (\mu=-3..3)$. The brackets of
$\frak{so}(5)$ over this basis are given in Table 2. According to
\cite{Me}, the Casimir operators of $\frak{so}(3)$ and
$\frak{so}(5)$ are given respectively by:

\begin{eqnarray*}
\fl
C_{21}=l_{0}^{2}+l_{1}l_{-1},\\
\fl
C_{2}=l_{0}^{2}+l_{1}l_{-1}-\frac{2}{5}\left(q_{3}q_{-3}+q_{1}q_{-1}\right)+\frac{1}{15}q_{2}q_{2}+q_{0}^{2},\\
\fl C_{4}=l_{0}^{3}q_{0}+\frac{1}{6}\left(  l_{-1}q_{1}-l_{1}q_{-1}+\frac{1}{2}%
l_{1}l_{-1}\right)  q_{0}^{2}+\frac{1}{6}\left(  q_{3}q_{-1}q_{-2}+q_{2}%
q_{1}q_{-3}+\frac{1}{3}q_{1}^{2}q_{-2}+\frac{1}{3}q_{2}q_{-1}^{2}\right)
q_{0}+\\
\fl -\frac{1}{3}\left(
\frac{1}{3}l_{-1}q_{-1}+\frac{1}{2}l_{0}q_{-2}+\frac
{2}{3}l_{1}q_{-3}\right)  q_{1}^{2}+\frac{1}{3}\left(  \frac{2}{3}l_{-1}%
q_{3}+\frac{1}{3}l_{1}q_{1}-\frac{1}{2}l_{0}q_{2}\right)
q_{-1}^{2}+\frac
{1}{4}\left(  l_{1}^{3}q_{-3}-l_{-1}^{3}q_{3}\right) \\
\fl \frac{1}{3}\left(
\frac{1}{20}q_{2}q_{-2}-q_{1}q_{-1}-3l_{-1}q_{1}+\frac
{7}{4}q_{0}^{2}+3l_{1}q_{-1}+\frac{1}{5}q_{3}q_{-3}\right)
l_{0}^{2}-\frac
{3}{100}q_{3}^{2}q_{-3}^{2}-\frac{q_{-2}q_{2}}%
{540}\left(  q_{1}q_{-1}+36q_{0}^{2}\right)\\
\fl +\frac{1}{12}\left(  q_{-1}^{2}-3l_{-1}q_{-1}+3l_{0}q_{-2}+q_{1}q_{-3}%
-q_{0}q_{-2}\right)  l_{1}^{2}+\frac{1}{12}\left(  3l_{0}q_{2}+q_{3}%
q_{-1}+q_{1}^{2}+3l_{1}q_{1}-q_{2}q_{0}\right)  l_{-1}^{2}\\
\fl +\frac{1}{3}\left(  -\frac{11}{20}l_{1}l_{-1}+l_{-1}q_{1}-\frac{3}{2}%
l_{0}q_{0}-l_{1}q_{-1}\right)  q_{-3}q_{3}+\frac{1}{6}\left(  \frac{1}%
{10}l_{1}l_{-1}-\frac{q_{-2}q_{2}}{6}l_{1}q_{-1}+\frac{2}{3}l_{0}q_{0}+\frac{1}{6}%
l_{-1}q_{1}\right) \\
\fl -\frac{1}{12}\left(  l_{1}l_{-1}-\frac{34}{3}l_{0}q_{0}\right)  q_{-1}%
q_{1}+\frac{1}{4}\left(  l_{1}q_{2}q_{-3}-\frac{1}{9}l_{-1}q_{2}q_{-1}%
+\frac{1}{9}l_{1}q_{1}q_{-2}-l_{-1}q_{3}q_{-2}\right)  q_{0}+\frac{q_{2}^{2}q_{-2}^{2}}{675}\\
\fl -\frac{1}{6}\left(  9l_{1}l_{-1}+l_{-1}q_{1}-l_{1}q_{-1}\right)  l_{0}%
q_{0}+\frac{1}{12}\left( \left(  q_{2}q_{-3}-q_{1}q_{-2}\right)
l_{1}l_{0}+\left(  -q_{3}q_{-2}+q_{2}q_{-1}\right)  l_{-1}l_{0}\right)-l_{0}q_{0}^{3}\\
\fl +\frac{1}{18}l_{0}q_{2}q_{1}q_{-3}-\frac{1}{36}\left(  q_{2}^{2}q_{-1}%
q_{-3}-l_{1}q_{3}q_{-2}^{2}+q_{3}q_{1}q_{-2}^{2}+l_{-1}q_{2}^{2}q_{-3}\right)
-\frac{1}{9}\left( q_{1}^{3}q_{-3}+q_{3}q_{-1}^{3}\right)
-\frac{5}{108}q_{1}^{2}q_{-1}^{2}\\
\fl +\frac{1}{5}\left(  \frac{7}{6}q_{1}%
q_{-1}-3q_{0}^{2}+\frac{1}{20}q_{2}q_{-2}\right)  q_{-3}q_{3}+\frac{1}{18}%
l_{0}q_{3}q_{-1}q_{-2}.%
\end{eqnarray*}

For this algebra, the transformations (\ref{TB}) defining the
contraction $\frak{g}$ are given by $L_{i}^{\prime}=L_{i},\;
Q_{\mu}^{\prime}=\frac{1}{t}Q_{\mu}$. The resulting algebra has an
Abelian radical of dimension seven, which implies that the
invariants will only depend on the $q_{\mu}$-variables \cite{C23}.
It is straightforward to verify that $\mathcal{N}(\frak{g})=4$,
and from the four Casimir operators, two can be obtained by
contracting the invariants $C_{2}$ and $C_{4}$ of $\frak{so}(5)$.
A basis of invariants of $\frak{g}$ is completed with two
operators $C_{6}^{\prime}$ and $C_{8}^{\prime}$ of degrees 6 and 8
respectively. Omitting $C_{8}$ because of its length, the explicit
form of the invariants $C_{2}^{\prime},C_{4}^{\prime}$ and
$C_{6}^{\prime}$ is as follows:

\begin{eqnarray*}
\fl
C_{2}^{\prime}=-\frac{2}{5}\left(q_{3}q_{-3}+q_{1}q_{-1}\right)+\frac{1}{15}q_{2}q_{2}+q_{0}^{2},\\
\fl C_{4}^{\prime}=\frac{1}{6}\left(  q_{3}q_{-1}q_{-2}+q_{2}q_{1}q_{-3}+\frac{1}%
{3}q_{1}^{2}q_{-2}+\frac{1}{3}q_{2}q_{-1}^{2}\right)  q_{0}
-\frac{1}{540}\left(  q_{1}q_{-1}%
+36q_{0}^{2}\right)  q_{-2}q_{2} \\
\fl -\frac{1}{36}\left(
q_{2}^{2}q_{-1}q_{-3}+q_{3}q_{1}q_{-2}^{2}\right)
-\frac{1}{9}\left(  q_{1}^{3}q_{-3}+q_{3}q_{-1}^{3}\right)
 +\frac{1}{5}\left(  \frac{7}{6}q_{1}q_{-1}-3q_{0}^{2}+\frac{1}{20}%
q_{2}q_{-2}\right)  q_{-3}q_{3}\\
\fl +\frac{q_{2}^{2}q_{-2}^{2}}{675}-\frac{5}%
{108}q_{1}^{2}q_{-1}^{2}-\frac{3}{100}q_{3}^{2}q_{-3}^{2},\\
\fl C_{6}^{\prime}=-729q_{0}^{6}-54q_{1}^{4}q_{-2}^{2}
+54q_{3}q_{-3}\left(  9q_{2}q_{0}^{2}q_{-2}+162q_{1}q_{0}^{2}q_{-1}%
-32q_{1}^{2}q_{-1}^{2}+6q_{2}q_{1}q_{-1}q_{-2}\right)  \\
 \fl +6q_{2}q_{-2}\left(  6q_{3}q_{-1}^{3}-10q_{1}^{2}q_{-1}^{2}+6q_{-3}q_{1}%
^{3}-63q_{1}q_{0}^{2}q_{-1}\right)  -162q_{0}^{2}\left(  q_{-2}^{2}q_{3}%
q_{1}+q_{2}^{2}q_{-3}q_{-1}\right)  \\
  \fl+54\left(  q_{0}^{2}\left(  27q_{3}^{2}q_{-3}^{2}-8q_{-3}q_{1}%
^{3}-8q_{3}q_{-1}^{3}-13q_{1}^{2}q_{-1}^{2}\right)
-q_{3}^{2}\left(
-q_{0}q_{-2}^{3}+q_{-1}^{2}q_{-2}^{2}\right)  -\left(  q_{1}^{2}q_{-3}%
^{2}+q_{-1}^{4}\right)  q_{2}^{2}\right)  \\
  \fl+972\left(  q_{0}^{3}\left(
q_{3}q_{-1}q_{-2}+q_{2}q_{1}q_{-3}\right)
-\left(  q_{3}^{2}q_{-1}q_{-2}q_{-3}+\left(  q_{2}q_{-1}^{2}q_{-3}+q_{2}%
q_{1}q_{-3}^{2}+q_{1}^{2}q_{-2}q_{-3}\right)  q_{3}\right)  q_{0}\right)  \\
 \fl+288q_{-1}q_{1}\left( q_{-3}q_{1}^{3}+q_{3}q_{-1}^{3}\right)
+90q_{-2}q_{2}\left(  q_{1}^{2}q_{-2}+q_{2}q_{-1}^{2}\right)  q_{0}%
+396q_{-1}q_{0}q_{1}\left(  q_{1}^{2}q_{-2}+q_{2}q_{-1}^{2}\right)  \\
 \fl +180q_{1}q_{-1}\left(  q_{-2}^{2}q_{3}q_{1}+q_{2}^{2}q_{-3}%
q_{-1}\right)  +864q_{-3}q_{3}\left(
q_{-3}q_{1}^{3}+q_{3}q_{-1}^{3}\right)+q_{2}^{3}q_{-2}^{3}-64q_{1}^{3}q_{-1}%
^{3}+q_{2}^{3}q_{0}q_{-3}^{2}\\
\fl -324q_{0}^{3}\left(  q_{1}^{2}q_{-2}+q_{2}q_{-1}^{2}\right)
-18q_{-2}q_{2}\left(
q_{-2}^{2}q_{3}q_{1}+q_{2}^{2}q_{-3}q_{-1}\right)
-756q_{0}q_{1}q_{-1}\left(  q_{3}q_{-1}q_{-2}+q_{2}q_{1}q_{-3}\right)\\
\fl +243\left(  6q_{1}q_{-1}-30q_{3}q_{-3}+q_{2}q_{-2}\right)
q_{0}^{4}-3q_{2}^{2}q_{-2}^{2}\left(
4q_{1}q_{-1}+9q_{0}^{2}\right)
\end{eqnarray*}

By inspection, we easily see that $C_{2}-C_{2}^{\prime}=C_{21}$,
therefore the set
$\left\{C_{2},C_{4},C_{21},C_{2}^{\prime},C_{4}^{\prime},C_{6}^{\prime}\right\}$
has at most rank five. Computing the Jacobian with respect to the
variables $\left\{q_{-3},q_{0},q_{1},l_{1},l_{0}\right\}$, we
prove that the rank is indeed five. We can therefore solve the
missing label problem. From the preceding functions we deduce that
a missing label operator is at least of order 4, thus reconfirming
the observation on the minimal degree of such an operator made in
\cite{Me}. This fourth order operator can be taken for example as
$\Phi_{1}=C_{4}-C_{4}^{\prime}+\frac{7}{12}C_{21}\left(C_{21}-C_{2}\right)$.
We point out that this choice does not coincide with that made in
\cite{Me}, where the simplest possible fourth order operator was
considered. A sixth degree missing label operator that commutes
with $\Phi_{1}$ can be taken as
$\Phi_{2}=C_{6}^{\prime}-13608C_{4}\left(C_{2}-C_{21}\right)+729\left(C_{2}^{2}-C_{21}^{3}\right)
+2187\left(C_{2}^{2}-C_{21}^{2}\right)$.

\begin{table}
\caption{\label{Tab1} $\frak{so}(5)$ brackets in a
$\frak{so}(3)=\left\{L_{0},L_{\pm 1}\right\}$ basis.}
\begin{indented}
\item[]\begin{tabular}[c]{@{}c|cccccccccc}%
$\left[  {}\right]  $ & $Q_{3}$ & $Q_{2}$ & $Q_{1}$ & $Q_{0}$ &
$Q_{-1}$ & $Q_{-2}$ & $Q_{-3}$\\\hline\mr
$L_{0}$ & $3Q_{3}$ &
$2Q_{2}$ & $Q_{1}$ & $0$ & $-Q_{-1}$ & $-2Q_{-2}$ &
$-3Q_{-3}$\\
$L_{1}$ & $0$ & $6Q_{3}$ & $Q_{2}$ & $2Q_{1}$ & $6Q_{0}$ &
$10Q_{-1}$ &
$Q_{-2}$\\
$L_{-1}$ & $Q_{2}$ & $10Q_{1}$ & $6Q_{0}$ & $2Q_{-1}$ & $Q_{-2}$ &
$6Q_{-3}$ &
$0$\\
$Q_{3}$ & $0$ & $0$ & $0$ & $Q_{3}$ & $Q_{2}$ & $10Q_{1}+15L_{1}$
&
$5Q_{0}-15L_{0}$\\
$Q_{2}$ &  & $0$ & $-6Q_{3}$ & $-Q_{2}$ & $-15L_{1}$ &
$30Q_{0}+60L_{0}$ &
$10Q_{-1}-15L_{-1}$\\
$Q_{1}$ &  &  & $0$ & $3L_{1}-Q_{1}$ & $-3L_{0}-3Q_{0}$ &
$15L_{-1}$ &
$Q_{-2}$\\
$Q_{0}$ &  &  &  & $0$ & $-Q_{-1}-3L_{-1}$ & $-Q_{-2}$ & $Q_{-3}$\\
$Q_{-1}$ &  &  &  &  & $0$ & $-6Q_{-3}$ & $0$\\
$Q_{-2}$ &  &  &  &  &  & $0$ & $0$ \\\hline\br
\end{tabular}
\end{indented}
\end{table}

\section{On the validity of the method}

The contraction method can constitute a practical procedure to
reduce to some extent the computations when we consider reduction
chains $\frak{s}\supset\frak{s}^{\prime}$ with more than three
missing labels, whenever the conditions of theorem 1 are
satisfied. For example, a solution for the general chains
$\frak{sp}(2N)\supset\frak{sp}(2N-2)\times\frak{u}(1)$ or
$\frak{sp}(2N)\supset\frak{sp}(2N-2)\times\frak{su}(2)$,
considered for the first time in \cite{Bi}, can be found by
analyzing the corresponding contractions.

\medskip

As has been pointed out when deriving formula (\ref{U3}), the
contraction method could fail if the contraction $\frak{g}$ has
``to few" invariants with respect to the number of necessary
labelling operators. Actually, this can happen for reductive
$\frak{s}^{\prime}$ and semisimple $\frak{s}$ if the following
numerical equality $\mathcal{N}(\frak{s})=\mathcal{N}(\frak{g})=n$
holds. Since in this case a fundamental system of invariants of
the contraction $\frak{g}$ can be obtained by appropriate
contraction of the Casimir operators of $\frak{s}$, the dependence
of the quadratic Casimir operator implies that we get at most
$n-1$ of the needed labelling operators. The remaining operator,
which must be computed explicitly, may however be determined in
some sense by the other operators, by means of the commutation
property it must satisfy. Although for this extreme case we don't
obtain a complete set by the contraction, it could also happen
that any degeneracy of practical interest can be resolved using
only the $n-1$ operators associated to the contraction. This
however requires a case by case inspection.

\medskip
The lowest dimensional reduction where the contraction produces an
insufficient number of labelling operators is the reduction
$G_{2}\supset \frak{su}(2)\times\frak{su}(2)$, where $G_{2}$ is
the exceptional Lie algebra of rank two. In this case, we have
$n=2$ missing labels, therefore four available operators. In
\cite{Hu1}, a pair of commuting operators of order six that solves
the missing label problem was found. The general form of commuting
operators remains however an unanswered question. Observe that
here,
$\mathcal{N}(G_{2})=\mathcal{N}(\left[\frak{su}(2)\right]^2)=2$
holds. In this case, the $G_{2}$ generators decompose as those of
the subalgebra and an eight dimensional irreducible representation
$R$ of $\frak{su}(2)\times\frak{su}(2)$, therefore the contraction
has the Levi decomposition
$\frak{g}=\left(\frak{su}(2)\times\frak{su}(2)\right)\overrightarrow{\oplus}_{R}8L_{1}$.
This algebra satisfies $\mathcal{N}(\frak{g})=2$. This means that
the invariants $C_{2}^{\prime}$ and $C_{6}^{\prime}$ of the
contraction algebra are obtained by limiting procedure from the
quadratic and hexic Casimir operators of $G_{2}$. Now the
quadratic operator is dependent on the operators of the same
degree of $G_{2}$ and the subalgebra. A routinary but cumbersome
computation shows that the function $C^{\prime}_{6}$ is
independent on the invariants of the algebra-subalgebra chain.
Therefore we arrive at a missing label operator $\Phi$ of degree
six, but a second independent operator cannot be constructed,
because there is no other independent higher order invariant in
the contraction. Taking into account the construction made in
\cite{Hu1}, this second operator must be either of degree six or
eight. Since both $G_{2}$ and $\frak{g}$ have at most one
(independent) invariant of order higher than two, the failure of
the contraction seems to be directly related to the order of the
required labelling operators.

\section{Conclusions}

We have shown that many physically relevant missing label problems
can be completely solved by using the properties of the reduction
chain $\frak{s}\supset\frak{s}^{\prime}$, by means of a Lie
algebra contraction associated to this reduction. Analyzing the
set of invariants of the three involved Lie algebras, suitable
commuting operators can be found that solve the missing label
problem. In this approach, the found operators inherit an
intrinsic meaning, namely as those terms of the Casimir operators
of $\frak{s}$ that get lost during contraction, up to some
combination of lower order invariants of $\frak{s}$ and
$\frak{s}^{\prime}$. We have recomputed some classical reductions
appearing in atomic and nuclear physics, obtaining complete
agreement with the result obtained by different authors and
techniques. Further we have furnished a natural explanation of the
order of these operators, which are directly related to the order
of the Casimir operators of the contracted Lie algebra. For the
special case of $n=m=0$, we have obtained a direct relation among
the invariants of $\frak{s}$ and $\frak{s}^{\prime}$ with those of
the contraction $\frak{g}$, which provides a new interpretation of
the contracted invariants.

\medskip

It seems natural that, whenever the reduction chain is
non-canonical and the reduction is not multiplicity free, the
information lost is somehow determined by the chain itself, and
not by a priori external techniques. In this sense, the missing
label operators which arise from the contraction $\frak{g}$ should
correspond to the natural choice of physical labelling operators,
as they are obtained using only the available information on the
algebra-subalgebra chain and their invariants. This suggests that
these could be the correct physical operators to be considered for
the labelling of states. An argument supporting this
interpretation is the equivalence of the contraction procedure
with the K-matrix method in the $\frak{su}(3)\supset\frak{so}(3)$
chain or the Wigner supermultiplet model. Whether the remaining
possibilities that arise from the general algebraic solution of
the missing label problem are physically more relevant than those
operators found by contraction, remains a question that should be
analyzed for any specific physical situation. All examples show
also that the affine contraction provides at most $n$ of the $2n$
available operators, thus induces a kind of partition in the set
of labelling operators. This suggests the existence of a certain
kind of hierarchy among these operators, as well as the fact that
some of them are not directly related to the properties of the
embedding of the subalgebra, and therefore not equivalent to
these. The next natural step is to analyze if the contraction
$\frak{g}$ can also be used to derive the eigenvalues of the
missing label operators.

\medskip

The failure of the proposed method for the special case
$\mathcal{N}(\frak{g})=\mathcal{N}(\frak{s})=n$ is essentially a
consequence of the existence of the quadratic Casimir operators
for reductive Lie algebras. In this situation, a similar
obstruction to obtain the sufficient number of labelling operators
will appear whenever the Lie algebra $\frak{s}$, the subalgebra
$\frak{s}^{\prime}$ and the contraction $\frak{g}$ have all a
Casimir operator of the same degree. In this case the invariant of
the contraction will be dependent, we thus loose one solution. How
to recover this operator without solving explicitly the system of
partial differential equations remains unanswered, as well as the
meaning of this lost solution. In spite of this incompleteness,
the method is still worthy to be applied, since often particular
degeneracies can be solved using less than the required labelling
operators \cite{Gr}.

\medskip

Finally, the contraction method, essentially reducing the
obtainment of missing label operators to the computation of
invariants of three Lie algebras, constitutes an appropriate class
of algebras to be tested with the geometrical method based on
moving frames, recently introduced in \cite{Bo,Bo2}, and tested
successfully for large types of algebras. In this frame, the
solving of differential equations is replaced by algebraic
systems, which can be often be solved in more effective manner.
This algorithm can be therefore applied more efficiently to obtain
a maximal number of independent invariants of the three Lie
algebras involved in the MLP. Further, this approach probably
allows to deduce some properties linking the corresponding
automorphism groups of these Lie algebras. Moreover, in the case
of non-reductive subalgebras, the geometric method provides
solutions avoiding complex realizations of the invariants,
therefore discarding supplementary complications that usually
arise from the analytical approach. Whether the method can be
implemented to compute directly the missing label operators, is a
problem that has still to be explored.

\section*{Acknowledgement}
The author would like to thank the Universidad Nacional de
Rosario, where part of this work was done, as well as H. de Guise
for helpful discussions and additional references. This work was
partially supported by the research grant MTM2006-09152 of the
Ministerio de Educaci\'on y Ciencia.

\newpage

\end{document}